\documentclass[%
 reprint,
nofootinbib,
 amsmath,amssymb,
 aps,
]{revtex4-2}

\usepackage[normalem]{ulem}
\usepackage[utf8]{inputenc}
\usepackage{graphicx}
\usepackage{xspace} 
\usepackage{dcolumn} 
\usepackage{bm} 
\usepackage[hidelinks=true, colorlinks=true, citecolor=teal]{hyperref}
\usepackage{xcolor} 
\usepackage{orcidlink}
\usepackage{multirow}
\usepackage{booktabs}
\usepackage[draft]{changes}

\newcommand{\dd}{{\rm d}}
\newcommand{\hn}{\boldsymbol{\hat{n}}}
\newcommand{\hk}{\boldsymbol{\hat{k}}}
\newcommand{\boldx}{\boldsymbol{x}}
\newcommand{\boldk}{\boldsymbol{k}}
\newcommand{\boldv}{\boldsymbol{v}}

\newcommand{\bv}{\boldsymbol{v}}
\newcommand{\bw}{\boldsymbol{w}}
\newcommand{\ns}{N_\text{side}}

\begin{document}

\title{Multipolar structure of the local expansion rate from incomplete sky data}

\author{João G. Vicente~\orcidlink{0009-0008-2912-8995}}
\email[]{jgabvicente.2000@uel.br}
\author{Thiago S. Pereira~\orcidlink{0000-0002-6479-364X}}
\email[]{tspereira@uel.br}
\author{Ricardo G. Rodrigues~\orcidlink{0000-0003-3824-5524}}
\email{ricardo.gonzatto11@uel.br}
\author{Sandro D. P. Vitenti~\orcidlink{0000-0002-4587-7178}}
\email{vitenti@uel.br}
\author{Vitoria M. Gomes~\orcidlink{0009-0001-1197-1991}}
\email{vitoria.menegon@uel.br}
\affiliation{Departamento de Física, Universidade Estadual de Londrina, Rod. Celso Garcia Cid, Km 380, 86057-970, Londrina, Paraná, Brazil}

\date{\today}

\begin{abstract}
    Using the Cosmicflows-4 data, we reconstruct the first multipolar moments of a general function describing the local expansion rate. In addition to the conventional harmonic approach, we employ a basis of symmetric and trace-free tensors to characterize the anisotropies of the expansion rate, allowing us to identify all directions associated with each of its multipoles. Focusing on objects in $z\in[0.01,0.05]$ in the CMB rest frame, we derive all $2\ell+1$ degrees of freedom in the multipoles $\ell=1,2$ and 3, which are split into one amplitude and $\ell$ unit vectors per multipole. To mitigate anisotropies induced by incomplete sky coverage, we introduce a pixel-based mask that removes poorly sampled pixels. The full-sky expansion rate is reconstructed using two independent approaches: a pseudo-inverse of the multipole-coupling kernel induced by the mask, and a maximum-likelihood estimate of the underlying full-sky field. These approaches are validated through simulations that explore different mask resolutions, cosmic variance and statistical noise. We find that the quadrupole and octupole amplitudes are consistent (at 95\% C.L.) with the expectations of a $\Lambda$CDM universe with linear and mild nonlinear perturbations, where the anisotropies of the expansion rate result from small peculiar velocities. The dipole amplitude, however, is inconsistent with $\Lambda$CDM predictions at 3.3$\sigma$, with a direction $(l, b) = (290^\circ, -4^\circ) \pm 5^\circ$ consistent with a bulk flow. This signal comes predominantly from sources in $z\in[0.03,0.05]$. Finally, we conduct alignment tests between the dipole, quadrupole, and octupole vectors. We confirm recent findings showing that the maxima of these multipoles are approximately located at $(290^\circ,-4^\circ)$. However, detailed tests using the complete vector structure of these multipoles reveal no evidence of alignments.
\end{abstract}

\maketitle

\section{Introduction}
Tests of the Cosmological Principle (CP)---the hypothesis that the Universe is statistically homogeneous and isotropic on sufficiently large scales---play a central role in modern cosmology. Under this assumption, the spacetime geometry of the Universe is described by the Friedmann–Lemaître–Robertson–Walker metric, which forms the basis of the standard \(\Lambda\)CDM cosmological model. The empirical validation of the CP is thus not merely a consistency check of a simplifying assumption but a fundamental test of the framework used to interpret most cosmological observations.

After the precise measurements of the cosmic microwave background (CMB) sky delivered by WMAP \cite{Hinshaw_2013,Komatsu_2011} and Planck \cite{aghanim2020planck, akrami2018planck}, and the increasing number of three-dimensional catalogs of the cosmos at low redshifts \cite{york2000sloan, jones20096df, giovanelli2005arecibo, zhang2024fast, said2025desi}, a number of cosmological tensions and statistical anomalies have emerged in the \(\Lambda\)CDM framework \cite{peebles2022anomalies, Abdalla:2022yfr,Aluri:2022hzs,CosmoVerseNetwork:2025alb}, motivating renewed interest in empirical tests of the CP. The disagreement in the measurements of \(H_0\) using CMB and supernova data is among the most robust tensions, with discrepancies reported as high as \(5.8\sigma\) \cite{riess2022comprehensive, breuval2024small}, prompting several studies on possible connections to large-scale violations of the CP \cite{Krishnan:2021dyb,Luongo:2021nqh,Krishnan:2021jmh,Camarena:2022iae,McConville:2023xav,Jia:2025prq} (see also \cite{Marra:2026ctg} for a recent review). Indications of large-angle statistical anomalies in the CMB date back to COBE~\cite{Hinshaw:1996ut}, and a plethora of explanations based on deviations from the CP have been proposed---see Refs.~\cite{Bull:2015stt,Abdalla:2022yfr,Aluri:2022hzs,CosmoVerseNetwork:2025alb} for comprehensive reviews. Although they have been reported with lower statistical significances in comparison to the Hubble tension, they are robust among different systematics, such as incomplete sky and foreground cleaning, and datasets, such as COBE, WMAP, and Planck.

At high redshifts, the CP has been very well tested. Starting with Big Bang nucleosynthesis, anisotropic expansion modifies the thermal history of the early Universe and alters the predicted abundances of light elements, such as deuterium and helium-4 \cite{Pitrou:2018cgg}. Observations of these abundances therefore place tight constraints on possible deviations from isotropy at redshifts of the order $z\sim10^{9-11}$ \cite{barrow1976light,barrow1977homogeneity,Thorne:1967zz}.

Moving towards lower redshifts, at the recombination epoch, an anisotropic background expansion alters the Hubble rate \cite{Ellis:1968vb}, the angular diameter distance  \cite{saunders1969observations,Fleury:2014rea}, and the coupling between different perturbative modes \cite{Pereira:2007yy,Gumrukcuoglu:2007bx,Pereira:2015pxa,Franco:2017pxt}, with observable consequences for the statistics of temperature and polarization CMB anisotropies at \(z\approx10^3\). This has allowed the use of Planck data to constrain the anisotropic rate of expansion at recombination to less than \(0.0001\%\) \cite{saadeh2016isotropic}. Post-recombination CMB anisotropies, such as the kinetic Sunyaev-Zel'dovich effect and spectral distortions, also have the power to constrain inhomogeneous models of the Universe \cite{Caldwell:2007yu,Zhang:2010fa}.

At low redshifts (\(z\approx1\)), a variety of tracers have been used to test the Cosmological Principle, including type Ia Supernovae \cite{kalus2013constraints,Bengaly:2015dza,Andrade:2018eta,colin2019evidence, soltis2019percent,rahman2022new,sorrenti2023dipole,sorrenti2025low,sah2025anisotropy}, galaxies \cite{Bengaly:2015xkw,Bengaly:2016amk,Bengaly:2017zlo,Rameez:2017euv}, quasars \cite{secrest2021test, secrest2022challenge, dam2023testing}, and x-rays \cite{Migkas:2020fza}. Future cosmic shear data are also expected to improve existing limits on a possible anisotropic dark energy \cite{Pereira:2015jya,Adam:2024kgs,Adam:2025zxh}. However, tests of spatial symmetries using low redshift data usually suffer from additional difficulties, such as small data samples and incomplete angular coverage. Moreover, they require a careful modeling of the peculiar velocity field, which is itself an important source of spatial anisotropies.

Large-scale coherent motions of galaxies provide a direct probe of cosmic isotropy in the nearby Universe~\cite{Tsagas:2025pxi}. In the \(\Lambda\)CDM model, peculiar velocities arise from gravitational instability and are expected to follow a statistically isotropic distribution whose variance can be predicted within linear perturbation theory. Measurements of bulk flows at small redshifts therefore offer an important consistency test of the \(\Lambda\)CDM framework \cite{Watkins:2008hf,Nusser:2011tu}. In particular, the amplitude and direction of the velocity dipole can be compared with theoretical expectations and with the kinematic dipole observed in the CMB, providing a complementary way to investigate possible preferred directions in the large-scale structure of the Universe \cite{Planck:2013rgv,Carrick:2015xza} --- a program known as the Ellis-Baldwin test \cite{Ellis:1984uka,Secrest:2025nbt}.

The Cosmicflows-4 (CF4) catalog \cite{Tully:2022rbj} is a compendium of velocities and distance moduli of galaxies and supernovae at very low redshifts (\(z\lesssim0.1\)) that possesses several desirable properties for tests of the CP. First, it contains a large number of objects with good coverage both in redshift and across the sky, allowing us to constrain measures of anisotropy with reasonably high signal-to-noise. Due to this, the impact of parts of the sky with poor or incomplete sampling, such as the Zone of Avoidance \cite{Kraan-Korteweg:2000dzx}, can be mitigated with the use of masks. Second, since CF4 combines different distance-measurement techniques (Tully-Fisher, Fundamental Plane, Supernovae Ia, Surface Brightness fluctuations, etc.), it allows for direct measurements of peculiar velocities. This renders isotropy tests with CF4 data particularly sensitive to a kinematic dipole, where the largest deviations from isotropy are expected to appear. Finally, CF4 data probes the linear and quasi-linear regime, which facilitates theoretical assessments through linear perturbations of the \(\Lambda\)CDM model.

In this work, we revisit existing tests of cosmic isotropy at low redshifts, introducing some methodological improvements along the way. Our test is based on the recently introduced \emph{expansion rate fluctuation} field \(\eta\) \cite{Kalbouneh:2022tfw}, defined, up to a redshift-dependent normalization, as the logarithm of \(z/d_L\), where the luminosity distance \(d_L\) is allowed to be an arbitrary function of the redshift \(z\) and line-of-sight \(\hn\). The dependence on the logarithm ensures that the expansion rate fluctuations have Gaussian errors, which are easier to model statistically. At the background level, the fluctuation field is zero if and only if the underlying space is homogeneous and isotropic, i.e., \(d_L=d_L(z)\), so that it works as a null indicator of spatial isotropy. At the perturbative level, the fluctuation field is proportional to small velocity perturbations, whose power spectrum we derive from first principles assuming a fiducial \(\Lambda\)CDM universe. The dipole, quadrupole, and octupole power spectrum amplitudes of the fluctuation field have been previously constrained using CF3 \cite{Kalbouneh:2022tfw} and CF4 \cite{Kalbouneh:2024szq,Salzano:2025ang} data. However, a reconstruction of all the angular degrees of freedom of each multipole, allowing, in particular, for a thorough assessment of possible alignments between them, has not been carried out so far.

We introduce two important methodological improvements over existing analyses. The first of them relies on the choice of a complete orthogonal basis for expanding the fluctuation field in terms of its multipoles: \(\eta = \sum_\ell \eta_\ell\). In addition to the usual spherical harmonic basis, where each multipole \(\eta_{\ell}\) is built from a combination of the spherical harmonics \(Y_{\ell m}\), we introduce an independent but equivalent decomposition where each multipole is written in terms of rank-\(\ell\) symmetric and trace-free tensors \cite{Thorne:1980ru}. While in the harmonic basis each multipole \(\eta_\ell\) is characterized by complex coefficients \(\eta_{\ell m}\), in the tensor basis the same multipole is completely characterized by one real amplitude, which we identify as the (square root of) the angular power spectrum of the fluctuation field, plus \(\ell\) unit vectors, known as multipole vectors~\cite{Copi:2003kt,Oliveira:2018sef}. The amplitude and multipole vectors contain the same \(2\ell+1\) real degrees of freedom encoded in the \(\eta_{\ell m}\), but with a few important differences. First, since the coefficients of the tensor decomposition are given by scalars and vectors, both of which are frame-independent, our tests are rotationally invariant by construction. Second, in the case of a Gaussian, homogeneous, and statistically isotropic universe, all dependence on the underlying cosmology resides in the angular power spectrum of each multipole \(\eta_\ell\), while the multipole vectors will only measure the complex phases of the field. Alignments among these vectors thus offer an important indication of deviations from the concordance statistical model.

Our second methodological improvement lies in the introduction of a mask throughout our pipeline. While objects in the CF4 catalog are fairly uniformly distributed in the interval \(0.01\leq z \leq 0.05\), their angular distribution is less uniform, most notably in the Zone of Avoidance. Since the multipolar reconstruction of the expansion rate depends on some pixelization scheme of the sphere, it is important to employ a mask where empty or poorly sampled pixels are removed, in order to minimize angular biases in our final results. We employ three masks, differing in their angular resolutions, which allows us to assess the robustness of our results. Next, we employ two independent methods to infer the underlying full-sky expansion rate from the masked one: the first through the pseudo-inverse of the multipole-coupling kernel introduced by the mask, and the second through a maximum likelihood estimation of the full-sky field. Working in the CMB rest frame, our results indicate the existence of a dipolar component larger than expected in the \(\Lambda\)CDM model, at a 3.3\(\sigma\) level, in the Galactic direction \((l,b)=(290^\circ,-4^\circ)\pm5^\circ\). These results provide an independent assessment of the CP in the low-redshift Universe.

This work is organized as follows. The theoretical basis for our analysis is derived in Sec.~\ref{sec:theory}, including a brief description of the expansion rate fluctuation field, of our masking procedure, and of the tensor basis used in our analysis. We present a brief overview of CF4 data in Sec.~\ref{sec:dataset}, and describe our methodology in Sec.~\ref{sec:methods}. Our main findings are presented in Sec.~\ref{sec:results}, where we also quantify their statistical significance, compare the derived angular power spectrum with \(\Lambda\)CDM predictions, and assess the degree of alignments between the multipole vectors. We conclude and present our perspectives in Sec.~\ref{sec:conclusions}.

\section{Theory}\label{sec:theory}

\subsection{The expansion rate fluctuation field}
A direct and model-independent way to look for anisotropies in the expansion rate is by allowing the luminosity distance, \(d_L\), to be a general function of the object's redshift $z$ and position $\hn=(\theta,\phi)$:
\begin{equation}
    d_L = d_L(z,\hn)\,.
\end{equation}
While we could proceed to reconstruct the multipoles of this function from the data, the quantity that is really measured, for which one assumes normally distributed errors, is the distance modulus \(\mu = 5\log(d_L)+25\).\footnote{Throughout this work, \(d_L\) will be measured in units of Mpc.} Given that \(d_L=cz/H_0\) at low redshifts in an expanding isotropic universe, this motivates the introduction of the following quantity \cite{Kalbouneh:2022tfw,Kalbouneh:2024szq}
\begin{equation}\label{eq:etadef}
    \eta(z,\hn) \equiv \log\left(\frac{z}{d_L(z,\hn)}\right) - \mathcal{M}(z)\,,
\end{equation}
as a null indicator of isotropy, since this quantity vanishes in the isotropic case, and its errors will follow the same distribution as those of the distance modulus. The monopole function \(\mathcal{M}(z)\) is introduced to ensure that \(\eta\) has no monopole, since we are not interested in fitting the average expansion rate. For arbitrary redshifts, $\eta$ can be written as a Taylor series in powers of $z$, with the expansion coefficients given by general cosmographic parameters \cite{Kalbouneh:2024szq}. For simplicity, we shall refer to \(\eta\) simply as the `fluctuation field'.

Even if the null hypothesis holds, the measured value of the fluctuation field does not have to be zero. Indeed, since the measured redshift \(z\) is related to the cosmological redshift \(z_c\) as
\begin{equation}
 z = z_c + v/c
\end{equation}
where \(v\) is the peculiar velocity field, fluctuations in \(v\) will lead to fluctuations in \(\eta\). Assuming that the null hypothesis holds and working in the CMB rest frame, these two are related by
\begin{equation}\label{eq:eta_v}
    \eta(z, \hn) = \frac{v(z, \hn)}{cz\ln 10 }\,,
\end{equation}
where we have used $v/c\ll z_c$. Therefore, the null hypothesis should test whether \(\eta\) is consistent with zero given \eqref{eq:eta_v} plus statistical and systematic uncertainties. We will use this expression in Sec.~\ref{sec:results} to derive the angular power spectrum of the \(\eta\) field in a \(\Lambda\)CDM universe. 

Clearly, the multipoles of Eq. \eqref{eq:etadef} cannot be directly reconstructed, since the luminosity distance is only sampled at a discrete set of points. Nevertheless, we can propose a discrete estimator of this field as
\begin{equation}\label{eq:etai}
    \hat{\eta}_i = \log(z_i) + 5 - \frac{\mu_i}{5} - \mathcal{M}_i\,,
\end{equation}
where $i$ refers to each source in the sample. Thanks to the large data set provided by the CF4 catalog, we can tessellate the observed sky into \(N_\text{pix}\) pixels, and replace \(\hat{\eta}_i\) by its average at each pixel \(p\):
\begin{equation}\label{eq:eta_p}
    \eta_p = \frac{\sum_{i\in p}\hat{\eta}_i/\sigma^2_i}{\sum_{i\in p}1/\sigma^2_i}\,,\qquad
    p=0,1,\dots,N_{\text{pix}}-1\,,
\end{equation}
where the sum is over all data points (e.g., galaxies) falling in pixel $p$, and \(\sigma_i\) is the error, which has two different contributions. First, there is the (Gaussian) error \(\sigma_{i,\mu}\) of the distance modulus $\mu_i$, provided by the catalog. Second, we also need to account for non-linearities in the peculiar velocities. We shall assume that the non-linear velocities are uncorrelated, with a typical variance given by $300~\text{km s}^{-1}/c$ \cite{sarkar2007bulk, scrimgeour20166df}. The contribution from a non-linear peculiar velocity is significantly more complex than a simple uncorrelated dispersion component, as considered here. However, since we are averaging galaxies' velocities over relatively large pixels, we expect correlations due to non-linearities to be subdominant for the resolutions we consider. We thus take \(\sigma_{i,z} = 300\text{km s}^{-1}/(cz_i\ln10)\), so that our error estimate is
\begin{equation}\label{eq:errors}
 \sigma_{i}^2 = \sigma_{i,\mu}^2 + \sigma_{i,z}^2\,.
\end{equation}

The set of \(\eta_p\) for all pixels forms a discretized map of the field \eqref{eq:etadef}, from which the multipolar coefficients can be estimated as
\begin{align}\label{eq:eta_lm}
    \eta_{\ell m} & = \frac{4\pi}{N_\text{pix}} \sum_{p=0}^{N_\text{pix}-1}{\eta}_p Y_{\ell m}(\theta_p,\phi_p)\,,\nonumber \\
                  & \approx \int \dd^2\hn\,{\eta}(\hn)Y^*_{\ell m}(\hn)\,,
\end{align}
where the last line becomes an equality in the limit of infinite data points. From now on, we shall omit any dependence of \(\eta\) on redshift, since it is implicit that we are averaging over all redshifts in a given pixel.

Given the data, we convert the pixel values \(\eta_p\) into multipolar coefficients \(\eta_{\ell m}\) using \texttt{healpy} \cite{Zonca2019}---a Python implementation of the equal-area and iso-latitude tessellation scheme introduced by \texttt{HEALPix}. It enables fast and accurate discretization of fields on the sphere, by dividing it into 12 base pixels, each of which can be recursively subdivided according to the resolution parameter \(\ns\), resulting in a total of \(12\,N_{\rm side}^2\) equal-area pixels. In what follows, we shall quote \(\ns\) when referring to different angular resolutions.

The reality condition of \(\eta(\hn)\) implies that \(\eta_{\ell m}^* = (-1)^m\eta_{\ell,-m}\), meaning that each multipole \(\ell\) is described by \(2\ell+1\) real degrees of freedom. In Ref. \cite{Kalbouneh:2022tfw}, one of these degrees of freedom was represented by the angular power spectrum of the fluctuation field, \(C_\ell = \langle\eta_{\ell m}\eta^*_{\ell m}\rangle\), which can be estimated from the data as
\begin{equation}\label{eq:cl_estimator}
    C_{\ell} = \frac{1}{2\ell+1}\sum_{m=-\ell}^{\ell}|\eta_{\ell m}|^2\,,
\end{equation}
while two other degrees of freedom were chosen to be the Galactic direction \((l,b)\) in the sky that maximizes the multipole \(\eta_\ell = \sum_m\eta_{\ell m}Y_{\ell m}\). For \(\ell=1\), these three numbers \(\{C_\ell, l, b\}\)  exhaust all the information contained in the dipole \(\eta_1\). 
However, these three numbers are insufficient to describe the full structure of \(\eta_{\ell}\) for \(\ell\geq2\), for which there are \(2\ell-2\) additional degrees of freedom. We fix this situation in Sec.~\ref{subsec:mvs}.

\subsection{Incomplete sky}
When estimating the harmonic coefficients \eqref{eq:eta_lm} from the data, a few systematic sources of anisotropies should be considered. First, the radial distribution of the CF4 sample reflects not only the underlying large-scale structure, but also inhomogeneities due to artificial selection effects. These inhomogeneities can be alleviated by choosing a sufficiently narrow bin of redshifts. Second, there is the issue of uneven angular distribution of data points. A strategy to mitigate this effect is to remove poorly sampled pixels with a mask. We will discuss our strategy for building such a mask in Sec.~\ref{sec:methods}. For the moment, let us stress that a full-sky implementation of the pixel-based estimator \eqref{eq:eta_p} depends on all pixels containing at least one data point, since the estimator would be ill-defined otherwise. In other words, a full-sky implementation can only be pursued for low-resolution maps. However, since low-resolution maps may still contain unevenly populated pixels, one strategy adopted has been to rotate the data across the sky in order to ensure that the least populated pixel contains the largest amount of objects~\cite{Kalbouneh:2022tfw}. Since \texttt{HEALPix} uses pixels with different shapes, this scheme may induce biases in the multipolar structure we are trying to estimate. Thus, a mask not only frees us from this procedure but also allows us to choose arbitrary pixelization schemes and consequently to reconstruct higher multipoles in \eqref{eq:eta_lm}.

In the presence of a mask \(W(\hn)\), the fluctuation field \(\eta(\hn)\) is replaced by \(W(\hn)\eta(\hn)\), with harmonic coefficients given by
\begin{equation}\label{eq:masked_coefficients}
    \tilde{\eta}_{\ell m} = \int \dd^2\boldsymbol{\hat{n}}\,W(\hn)\eta(\boldsymbol{\hat{n}})  Y_{\ell m}^{*}(\boldsymbol{\hat{n}})\,.
\end{equation}
If we now expand both \(W(\hn)\) and \(\eta(\hn)\) in terms of spherical harmonics, we find that the masked and full-sky harmonic coefficients of the fluctuation field are related as \cite{hivon2002master}
\begin{equation}
    \label{eq:masked_eta_lm}
    \tilde{\eta}_{\ell m} = \sum_{\ell', m'}K_{\ell m \ell' m'}\eta_{\ell'm'},
\end{equation}
where the multipole-coupling kernel $K_{\ell m \ell' m'}$ is defined as
\begin{multline}
    \label{eq:kernel_def}
    K_{\ell m \ell' m'} \equiv \sum_{L,M}(-1)^{M}\sqrt{\frac{(2\ell + 1)(2\ell' + 1)(2L + 1)}{4\pi}}\\
    \times w_{L M}
    \begin{pmatrix}
        \ell & \ell' & L \\
        0    & 0     & 0
    \end{pmatrix}
    \begin{pmatrix}
        \ell & \ell' & L \\
        -m   & m'    & M
    \end{pmatrix}\,.
\end{multline}
Here, \(w_{L M}\) are the harmonic coefficients of the mask \(W(\boldsymbol{\hat{n}})\), and the \(2\times3\) matrices represent angular momentum couplings, known as Wigner-3j symbols \cite{edmonds1996angular}.

\begin{figure}
    \includegraphics[width=\linewidth]{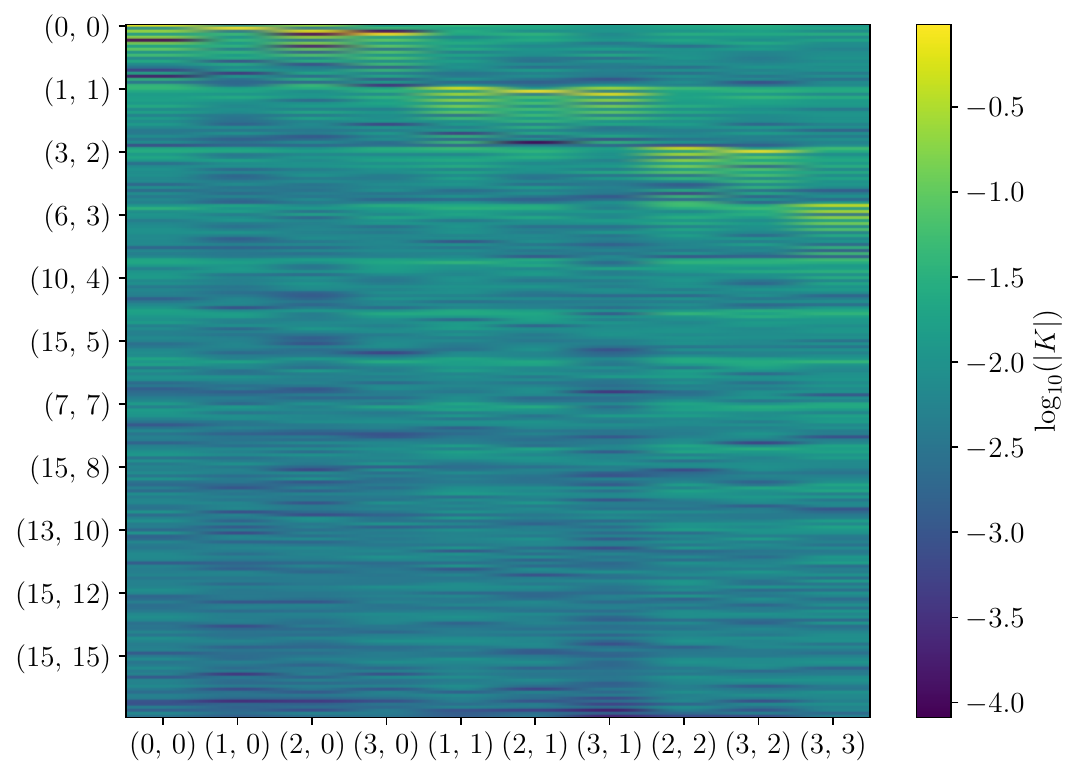}
    \caption{Absolute value of the multipole-coupling kernel $K_{\ell m \ell' m'}$, in logarithmic scale, for a typical mask used in this work. See Sec.~\ref{sec:methods} for details.}
    \label{fig:kernel}
\end{figure}

The Wigner-3j symbols enforce a triangular inequality between the multipoles \(\ell\), \(\ell'\) and \(L\), so that the main effect of the kernel is to couple the multipole moments of the full-sky field (\(\eta_{\ell m}\)) with those of the mask (\(w_{LM}\)). In other words, the masked coefficients are non-local in harmonic space, meaning that any multipole \(\ell\) of the masked field receives contributions from all multipoles \(\ell'\) of the full sky field obeying \(|L-\ell|\leq \ell'\leq L+\ell\). In practice, we shall be dealing with large-angle masks for which the coefficients \(w_{LM}\) quickly approach zero as \(L\) grows, so that the convergence of Eq. \eqref{eq:masked_eta_lm} is not an issue. Figure~\ref{fig:kernel} shows a portion of a typical kernel that we shall consider in this work.

Of course, since we are interested in the intrinsic anisotropies of the full-sky field \(\eta(\hn)\), rather than the systematic anisotropies of the field \(\tilde{\eta}(\hn)\), some inversion scheme is required. This is complicated by the fact that the non-zero entries of the kernel \eqref{eq:kernel_def} are not, in general, represented by a square matrix. Furthermore, because the mask removes portions of the data, an exact inversion is not possible. Thus, in Sec.~\ref{sec:methods} we discuss two approximations used to reconstruct the full sky coefficients \(\eta_{\ell m}\) from the masked ones. The first is through a pseudo-inverse of Eq.~\eqref{eq:masked_eta_lm}, and the second through a Bayesian estimate of the underlying full sky field.

\subsection{Symmetric and trace-free tensors}\label{subsec:mvs}
To analyze the angular structure of cosmological observables such as the fluctuation field \(\eta\), one needs a complete and orthonormal basis of functions on the sphere. The decomposition of observables in terms of spherical harmonics is widely employed in cosmology, from large-scale structure studies to the treatment of CMB anisotropies. It is ideally suited for linear problems, where different multipole moments can be studied independently. Nonetheless, this choice is ultimately guided by convenience, and other bases might be equally or more convenient depending on the problem at hand.

An equivalent spectral decomposition for functions on the sphere can be constructed in terms of totally symmetric and trace-free (STF) tensors~\cite{Thorne:1980ru,Copi:2003kt}. In this basis, the fluctuation field is decomposed as\footnote{Recall that, by the definition \eqref{eq:etadef}, \(\eta(\hn)\) has no monopole.}
\begin{equation}\label{eq:stf-decomp}
    \eta(\hn) = \eta^{i_1}n_{i_1} + \eta^{i_1 i_2}n_{i_1}n_{i_2} + \eta^{i_1 i_2 i_3}n_{i_1}n_{i_2}n_{i_3} + \cdots\,,
\end{equation}
where the indices run from 1 to 3, and sums are implied over repeated indices. Here, \(\hn=(\sin\theta\cos\phi,\sin\theta\sin\phi,\cos\theta)\) represents the angular dependence of the field, while the tensors $\eta^{i_1\cdots i_\ell}$ encode information about its multipolar content. The equivalence between \eqref{eq:stf-decomp} and the usual harmonic decomposition can be verified by noting that any spherical harmonic can be written as
\begin{equation}
    Y_{\ell m}(\hn) = {\cal Y}^{i_1\cdots i_\ell}_{\ell m}n_{i_1}\cdots n_{i_\ell}
\end{equation}
where \({\cal Y}^{i_1 i_2\cdots i_\ell}_{\ell m}\) are rank-\(\ell\) STF tensors in the upper indices, given by combinations of Kronecker deltas \cite{Thorne:1980ru}. It then follows from \(\eta=\sum_{\ell m}\eta_{\ell m}Y_{\ell m}\) that
\begin{equation}
    \eta^{i_1\cdots i_\ell}=\sum_{m=-\ell}^{\ell}\eta_{\ell m}{\cal Y}^{i_1 i_2\cdots i_\ell}_{\ell m}
\end{equation}
are also STF tensors.

An interesting property of this expansion lies in the fact that, since STF tensors of rank-\(\ell\) in three dimensions contain \(2\ell+1\) real degrees of freedom, they can be put into one-to-one correspondence with one constant \(\lambda_\ell\) and \(\ell\) unit vectors \(\bv_{(\alpha,\ell)}\), with \(\alpha=1,2,\cdots\ell\) \cite{Rodrigues:2024pkh}. In other words,
\begin{align}
    \eta^{i_1}         & = \lambda_1 v^{i_1}_{(1,1)}\,,\quad                                            \\
    \eta^{i_1 i_2}     & = \lambda_2 v^{\langle i_1}_{(1,2)}v^{i_2\rangle}_{(2,2)}\,,\quad              \\
    \eta^{i_1 i_2 i_3} & = \lambda_3 v^{\langle i_1 }_{(1,3)} v^{i_2}_{(2,3)} v^{i_3\rangle}_{(3,3)}\,,
\end{align}
and so on, where the brackets \(\langle\cdots\rangle\) denote a fully symmetric and trace-free combination of the enclosed indices. Note that the constant \(\lambda_\ell\) and the vectors \(\bv_{(\alpha,\ell)}\) are determined only up to an overall sign, since \((\lambda_\ell,\bv_{(\alpha,\ell)})\rightarrow (-\lambda_\ell,-\bv_{(\alpha,\ell)})\) leaves \eqref{eq:stf-decomp} unchanged for any \(\alpha\). Strictly speaking, these are not vectors, but lines piercing the sphere at a pair of antipodal points.\footnote{To be precise, they are elements of the projective space \(\mathbb{RP}^2\).} This antipodal symmetry is the STF version of the reality condition of the field \(\eta\) which, in the harmonic expansion, is ensured by \(\eta^*_{\ell m}=(-1)^m \eta_{\ell,-m}\). The vectors \(\bv_{(\alpha,\ell)}\) are known as multipole vectors, and were originally introduced by J. C. Maxwell \cite{maxwell1873treatise} to represent the multipole moments of charge distributions. They were reintroduced into cosmology as a diagnostic tool to test the isotropy and Gaussianity of CMB \cite{Copi:2003kt}, and have since been widely used in the study of CMB anomalies \cite{Abramo:2009fe,Pinkwart:2018nkc,Oliveira:2018sef}. An efficient code converting the harmonic coefficients of any function into its multipole vectors, and dubbed \texttt{PolyMV}, was introduced in~\cite{oliveira2020cmb}. For any multipole \(\ell\), \texttt{PolyMV} returns \(2\ell\) points on the unit sphere, corresponding to the vectors and their antipodes, in no particular order.

The STF and harmonic bases are equivalent, and thus the sets \(\{\lambda_\ell, \bv_{(\alpha,\ell)}\}\) and \(\{\eta_{\ell m}\}\) represent the same \(2\ell+1\) degrees of freedom of a given real function on the sphere. However, the STF basis has a few important properties which, depending on the circumstances, make them preferable to the harmonic basis. First, under SO(3) rotations, the vectors \(\bv_{(\alpha,\ell)}\) rotate rigidly with the data, so that their components do not get mixed. This is in sharp contrast to the coefficients \(\eta_{\ell m}\) which, under rotations, transform as \(\eta_{\ell m}=\sum_m D^\ell_{m m'}\eta_{\ell m'}\), where \(D^{\ell}_{m' m}\) are Wigner rotation matrices \cite{edmonds1996angular}. This property is particularly useful: since the data alone determine the vectors, without any reference to particular reference frames, our tests are less prone to biases introduced by particular choices of coordinates. Second, in the hypothesis of a Gaussian, homogeneous and isotropic universe, all the dependence on the cosmological parameters---which in the harmonic basis is encoded in the standard deviation \(\sqrt{C_\ell}\) of the harmonic coefficients \(\eta_{\ell m}\)---is in the STF basis carried by the constants \(\lambda_\ell\), and not by the vectors \(\bv_{(\alpha,\ell)}\)~\cite{Oliveira:2018sef}. Thus, in the concordance statistical framework, the constants \(\lambda_\ell\propto\sqrt{C_\ell}\) contain all the information about the cosmology, while the vectors \(\bv_{(\alpha,\ell)}\) only reflect the random phases of the stochastic field that they describe. Their distribution function, as well as any indicator of alignments between these vectors, thus offers an important null-test of the standard statistical hypothesis.

When dealing with random points on the sphere, such as the multipole vectors of the fluctuation field, or any other vectors derived from it, care must be taken regarding the fact that these are random variables on a compact space. Consequently, the computation of statistical moments, such as their mean and standard deviation, must properly account for the curvature of the sphere. The Fréchet variance is a direct generalization to metric spaces of the usual notion of variance \cite{frechet1948elements}. For a set with \(N\) realizations of the vector \(\bv_{(\alpha,\ell)}\), we define the (sample) Fréchet variance as
\begin{equation}\label{eq:frechet-var}
    \Psi(\alpha,\ell) = \frac{1}{N}\sum_{i=1}^{N} [\arccos(\bar{\bv}_{(\alpha,\ell)}\cdot\boldsymbol{v}^i_{(\alpha,\ell)})]^2
\end{equation}
where the vector
\begin{equation}\label{eq:frechet-mean}
    \bar{\bv}_{(\alpha,\ell)} = \text{argmin}\,\Psi(\alpha,\ell)
\end{equation}
defines the Fréchet mean. We will use these definitions to extract the mean values and error bars of the vectors associated with the CF4 data in Sec.~\ref{sec:results}. Our \(1\sigma\) estimates of the error associated with the vectors \(\bv_{(\alpha,\ell)}\) correspond to \(\sqrt{\Psi(\alpha,\ell)}\).

In this work, we use \texttt{PolyMV} to compute the vectors \(\bv_{(\alpha,\ell)}\) from the (reconstructed) harmonic coefficients \(\eta_{\ell m}\). Since the constants \(\lambda_\ell\) contain the same information as the angular power spectrum \(C_\ell\), we quote the latter, as it is more easily obtained from \(\eta_{\ell m}\) using \eqref{eq:cl_estimator}.

\section{Dataset}\label{sec:dataset}

Reconstructing the degrees of freedom of each multipole of the fluctuation field requires a sufficiently large data set with broad sky-coverage. Cosmicflows-4 (CF4) \cite{Tully:2022rbj} provides the ideal catalog for this task. It consists of the largest compiled catalog of galaxy distances and peculiar velocities to date, and has been used to map the dynamics of the local universe \cite{valade2024identification}, as well as to study its large scale structure \cite{courtois2023gravity}. CF4 contains 55,877 individual galaxy distances, measured from multiple and independent methods, the largest fraction of them resulting from the Tully-Fisher (TF) and Fundamental Plane (FP) relations, with errors of about $\sim 20 \%$, and the remaining consisting of several other methods, including type Ia Supernovae, Surface Brightness Fluctuations, among others.

\begin{figure}
    \centering
    \includegraphics[width=\linewidth]{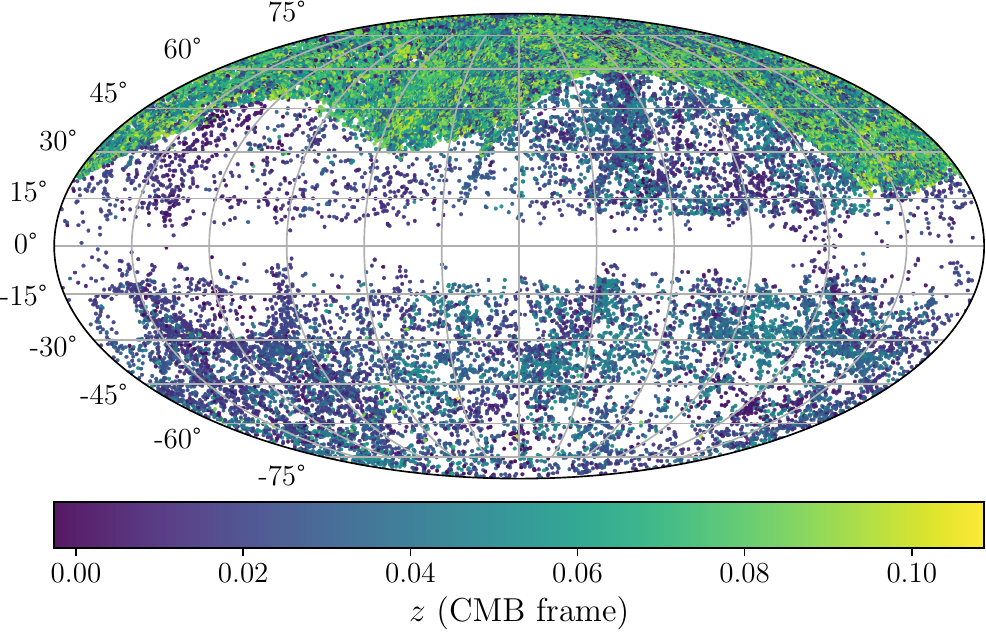}
    \caption{Spatial distribution of all 55,877 objects in the CF4 catalog. Besides the galactic plane, the data is distributed quite inhomogeneously in the northern hemisphere. Such inhomogeneities can be attenuated by choosing a narrower redshift bin (Figure \ref{fig:cf4_data}) and a mask (Figure \ref{fig:masks}).}
    \label{fig:cf4_all_data}
\end{figure}

Despite being a large catalog, CF4 data are distributed rather anisotropically, most notably in the Zone of Avoidance, the region around the galactic plane where interstellar dust and stars hinder observations of extragalactic objects---see Figure~\ref{fig:cf4_all_data}. As we have argued, these anisotropies could introduce spurious directions in our analysis, an effect that can be mitigated by introducing a mask. Moreover, the redshift distribution across the sky is highly anisotropic, and galaxies with redshifts \(\gtrsim 0.05\) are mainly located in the northern hemisphere. In order to avoid these anisotropies, we have chosen to work with redshifts in the range $z\in [0.01,0.05]$. This redshift interval reduces the original sample from $55,877$ to $24,646$ data points, $30\%$ of which are Fundamental Plane distances from Sloan Digital Sky Survey (SDSS) spectroscopic and photometric observations \cite{howlett2022sloan}, $30\%$ being Tully-Fisher distances from the Cosmicflows-4 Tully Fisher Catalog \cite{kourkchi2020cosmicflows}, and $25\%$ being Fundamental Plane distances from the 6-degree Field Galaxy Survey (6dFGS) \cite{springob20146df}.

\begin{figure}
    \includegraphics[width=\linewidth]{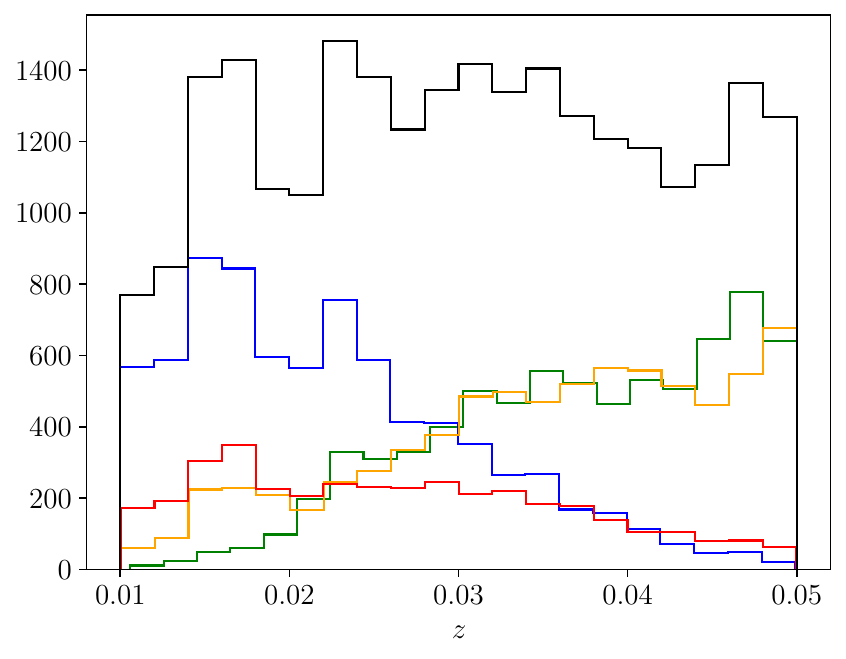}
    \caption{Redshift distributions of sources in the range $[0.01,0.05]$. Colors indicate the subsets from which each galaxy was sampled. Green: SDSS; Blue: CF4-TF; Orange: 6dFGS; Red: Others. The black histogram is the total redshift distribution in this interval.}
    \label{fig:cf4_data}
\end{figure}

Figure \ref{fig:cf4_data} shows the distribution of object counts as a function of redshift for the selected interval. Although the redshifts obtained from different distance methods are not uniformly distributed, the total distribution (black line in Fig. \ref{fig:cf4_data}) is better approximated by a uniform distribution. We stress that the choice of this redshift range is a limitation imposed by the data set, and not a limitation of our method.

\section{Methods}\label{sec:methods}

\subsection{Masks}\label{subsec:masks}

The first step in our implementation is to convert the discrete function \eqref{eq:etai} into a \texttt{healpy} map, given by \eqref{eq:eta_p}. Since the convergence of this map depends on a minimum number of points per pixel, and given that we have at most 24,646 galaxies at our disposal (i.e., before masking), our implementation is limited to a maximum number of pixels or, equivalently, to a maximum \(\ns\). In order to test the robustness of our method under different masks and total number of samples, we have chosen three different resolutions. After several tests, we concluded that the values \(\ns=4\), 8, and 16 are a good compromise between resolution, convergence, and fraction of sky coverage.

\begin{figure*}
    \includegraphics[width=\linewidth]{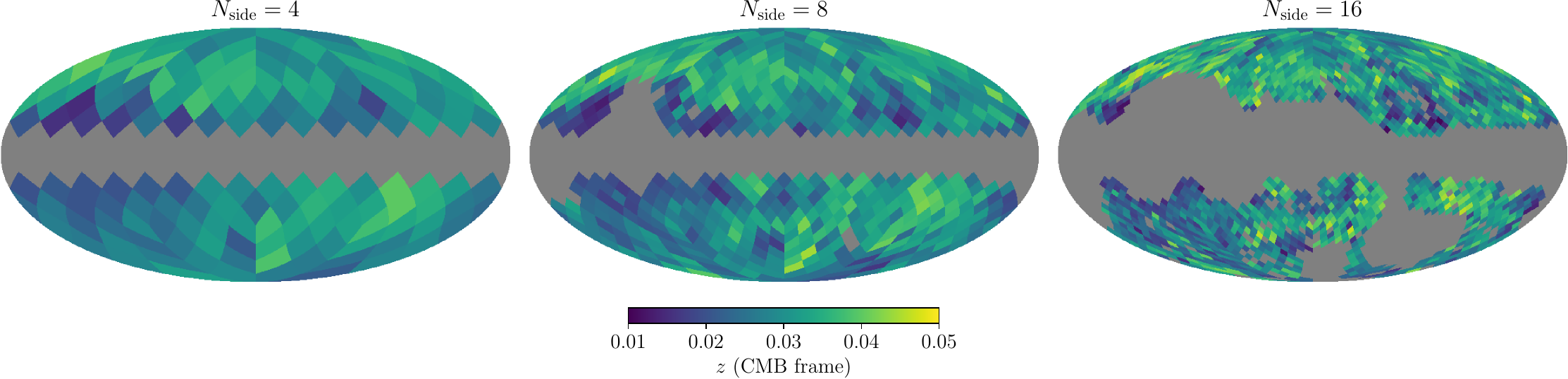}
    \caption{Masks adopted in this work, for different sky resolutions. For \(\ns=4\), the mask has azimuthal symmetry. The colors represent the mean redshift of galaxies in a given pixel, and for \(z\in[0.01,0.05]\)}
    \label{fig:masks}
\end{figure*}

Once the resolution is fixed, we need to define our mask. In order to minimize the variance of object counts per pixel, and thus ensure a more homogeneous sampling, we have developed some criteria for when a pixel should be masked. Thus, a pixel is masked if:
\begin{itemize}
    \item it has fewer than some minimum number of objects. This minimum depends on $\ns$, and we have adopted a minimum of 14, 5, and 3, respectively, for $\ns=4,8,16$;
    \item it lies within a constant-width strip of $10^\circ$ around the equator. This is slightly larger than the Zone of Avoidance region defined by 6dFGS~\cite{springob20146df};
    \item more than 50\% of its neighbors are masked. This procedure is repeated until convergence, and ensures that data is distributed more homogeneously over unmasked pixels.
\end{itemize}

Figure~\ref{fig:masks} shows the masks obtained with the above criteria for the three resolutions that we have adopted, together with the mean redshift in the unmasked region. Each mask then tells us where to evaluate Eq.~\eqref{eq:eta_p} and, consequently, Eq.~\eqref{eq:etai}. However, since the quantity \(\mathcal{M}_i\) is computed a posteriori to ensure that the full-sky fluctuation field has no monopole (i.e., $\eta_{00} = 0$), the quantity that is directly computed from the data is
\begin{equation}
    \hat{\beta}_i \equiv \log(z_i) - \frac{\mu_i}{5} + 5\,,
\end{equation}
so that $\hat{\eta}_i = \hat{\beta}_i - \mathcal{M}_i.$ We compute the average of this quantity on each pixel following the same procedure as \eqref{eq:eta_p}, leading to $\beta_p$. The inclusion of a mask then gives us $\tilde{\beta}_p = W_p\beta_{p}$ at each pixel $p$, where $W_p$, the mask in pixel space, is implemented as a boolean array. We then use \texttt{healpy} to compute $\tilde{\beta}_{\ell m}$, which are related to the harmonic coefficients of the masked map $\tilde{\eta}$ as
\begin{equation}\label{eq:etatilde}
    \tilde{\eta}_{\ell m} = \tilde{\beta}_{\ell m} - \mathcal{M}w_{\ell m}\,.
\end{equation}

Recall that our goal is to obtain the full-sky \(\eta_{\ell m}\) from the masked \(\tilde{\eta}_{\ell m}\), which requires evaluating the kernel \(K_{\ell m \ell' m'}\) appearing in Eq.~\eqref{eq:masked_eta_lm}. Given $\ell$ and $\ell'$, the sum in Eq.~\eqref{eq:kernel_def} is bounded in the interval \(|\ell - \ell'| \leq L \leq \ell + \ell'\) due to the triangular inequality of the Wigner-3j symbols. However, due to the finite resolution imposed by $\ns$, we can only compute $w_{L M}$ up to $3\ns - 1$. As such, this limits the maximum values of $\ell$, $\ell'$, and $L$. Obviously, it is impossible to reconstruct the exact full sky $\eta_{\ell m}$ when a mask is applied, simply because there is no information about this field in the masked region. Therefore, since we are interested in the first few multipoles of the full-sky map, we choose $\ell'_{\text{max}} = 3$, which limits $\ell_{\text{max}} = 3\ns - 4$.

Given the kernel computed as above, and equation \eqref{eq:etatilde}, we can now estimate $\eta_{\ell m}$. Note that Eq.~\eqref{eq:masked_eta_lm} forms an overdetermined system of equations for \(\ns=4\), 8 and 16, with $(\ell_{\text{max}} + 1)^2 = (3\ns - 3)^2$ equations for $(\ell'_{\text{max}} + 1)^2 = 16$ variables. Since, in general, there are no solutions to this problem, we employ two independent methods to approximate $\eta_{\ell m}$, which we now detail.

\subsubsection{Pseudo-inverse}
A possible solution to the problem above is to construct a pseudo-inverse matrix for the kernel $K_{\ell m \ell' m'}$. This method is very efficient since, for a given mask, the pseudo-inverse matrix needs to be computed only once.

Let \(K^+_{\ell m \ell' m'}\) be the pseudo-inverse of $K_{\ell' m' \ell'' m''}$, such that the absolute value of the difference
\begin{equation}\label{eq:pseudoinverse}
    \bigg|\sum_{\ell',m'}K^+_{\ell m \ell' m'} K_{\ell' m' \ell'' m''} - \delta_{\ell \ell''}\delta_{m m''}\bigg|
\end{equation}
is minimal. The full-sky coefficients $\eta_{\ell m}$ in Eq.~\eqref{eq:masked_eta_lm} can then be estimated as
\begin{equation}\label{eq:eta-reconstructed}
    \eta_{\ell m} \approx \sum_{\ell', m'}K^+_{\ell m \ell' m'} (\tilde{\eta}_{\ell'm'}-\mathcal{M}w_{\ell'm'})\,.
\end{equation}
Since \(\eta_{00} = 0\) by definition, this fixes \(\mathcal{M}\) as
\begin{equation}
    \mathcal{M} = \frac{\sum_{\ell'm'}K^+_{00 \ell'm'} \tilde{\beta}_{\ell'm'}}{\sum_{\ell'm'}K^+_{00 \ell'm'} w_{\ell'm'}}.
\end{equation}
Using \texttt{SciPy}'s default pseudo-inverse routine to obtain $K^+_{\ell m \ell'm'}$, we have checked that the largest value of Eq.~\eqref{eq:pseudoinverse} is at most of order $10^{-14}$. 

In order to estimate the statistical error in the reconstruction of the full-sky coefficients, we use Monte Carlo simulations to generate new datasets. For each CF4 galaxy \(g\) in the range \(z\in[0.01,0.05]\), we keep its direction and redshift fixed, and generate new distance moduli by sampling from the distribution \(\mathcal{N}(\mu_g, \sigma_g^2)\), where \(\mu_g\) and \(\sigma_g\) are the galaxy's reported distance modulus and standard deviation, respectively. By repeating the masking procedure described above for each of these simulations, we can estimate the statistical error in the reconstruction of \eqref{eq:eta-reconstructed}.

\subsubsection{MCMC}\label{subsec:MCMC}

To check the robustness of our results, we perform an alternative inference to the pseudo-inverse method. Since pseudo-inverse matrices offer a least-squares solution to an overdetermined system of equations, we extend this approach to include the statistical errors in the inversion procedure. That is, we introduce the \(\chi^2\) statistic:
\begin{equation}
    \chi^2 = \sum_{\ell m}\sum_{\ell' m'}v^*_{\ell m} \text{Cov}^{-1}_{\ell m \ell' m'}v_{\ell'm'}
\end{equation}
where
\begin{equation}
    v_{\ell m} = \beta_{\ell m} - \mathcal{M}w_{\ell m} - \sum_{\ell'm'}K_{\ell m\ell'm'}\eta_{\ell'm'}
\end{equation}
and look for $\mathcal{M}$ and $\eta_{\ell m}$ that minimize \(\chi^2\).

To estimate the covariance matrix, we perform Monte Carlo simulations similar to those described in the previous section, computing $\tilde{\beta}_{\ell m}$ for $N = 1000$ simulated datasets and then estimating the covariance as follows:
\begin{equation}
    \text{Cov}_{\ell m \ell'm'} = \frac{1}{N}\sum_{i = 1}^N (\tilde{\beta}_{\ell m}^i - \tilde{\beta}_{\ell m})^* (\tilde{\beta}_{\ell'm'}^i - \tilde{\beta}_{\ell'm'}),
\end{equation}
where $\tilde{\beta}_{\ell m}^i$ are the inferred masked multipoles for the \(i\)th simulation. The inverse covariance is then computed from the unbiased estimator given in \cite{Hartlap:2006kj}. We use the package \texttt{emcee} \cite{Foreman_Mackey_2013} to minimize the \(\chi^2\) and to obtain $\mathcal{M}$ and $\eta_{\ell m}$.

\subsection{Comparison with $\Lambda$CDM}\label{subsec:lcdm}

Thanks to the relation between the expansion rate fluctuation and the peculiar velocity fields (see Eq.~\eqref{eq:eta_v}), we can use linear perturbation theory to derive the theoretical angular power spectrum for the former. We show in Appendix \ref{sec:Cl_lcdm} that the angular spectrum is given by:
\begin{equation}\label{eq:Cl_lcdm}
    C_\ell=
    \frac{2f^2}{\pi(\ln 10)^2\Delta\chi}
    \int \dd k\,P_m(k)
    \left(\int_{\chi_{\text{min}}}^{\chi_{\text{max}}}
    \dd\chi\,\frac{j'_\ell(k\chi)}{\chi}\right)^2\,,
\end{equation}
where \(P_m(k)\) is the linear matter power spectrum at scale \(k\) and comoving distance \(\chi\), a prime in \(j_\ell\) denotes a derivative with respect to its argument and \(\Delta\chi=\chi_\text{max}-\chi_\text{min}\) accounts for sources in a fixed comoving distance bin. We use \texttt{CLASS} \cite{Diego_Blas_2011} and its default values for cosmological parameters to compute the linear \(P_m(k)\).\footnote{We have checked that the results are essentially the same if we instead use \texttt{CLASS}'s default halofit routine to compute \(P_m(k)\).}

In order to compare the measured \(C_\ell\) to Eq.~\eqref{eq:Cl_lcdm}, we also need to estimate the statistical and systematic uncertainties. This is done by generating mock catalogs, where we keep the angular positions of CF4 galaxies fixed, and generate pairs \((z_i,\mu_i)\) of `observed' redshift and distance modulus for each new simulation. These redshifts and distance moduli are generated as follows:

\begin{enumerate}
    \item For each galaxy \(i\) in a pixel \(p\), we define its ``cosmological'' redshift to be the mean redshift of all galaxies in pixel \(p\).
    \item From Eq.~\eqref{eq:Cl_lcdm}, we use \texttt{healpy} to generate a new realization of the map \(\eta(\hn)\). For this step, it is enough to work with \(\ell_\text{max}=6\), since the theoretical power spectrum falls steeply with increasing \(\ell\), as can be easily checked. From this map we compute the peculiar velocity through Eq.~\eqref{eq:eta_v}, using the CF4 angular positions and the cosmological redshifts computed in step 1. This step produces \(v_{i,\text{linear}}\).

    \item As discussed above Eq.~\eqref{eq:errors}, to this linear velocity we add a non-linear velocity component drawn from a normal distribution $\mathcal{N}(0, \sigma_z^2)$, where $\sigma_z$ is a 1D velocity dispersion usually taken to be $300~\text{km s}^{-1}/c$. To check the impact of this step, we have increased $\sigma_z$ up to $500~\text{km s}^{-1}/c$, finding that it does not alter the dipole intensity but slightly increases the chances of higher octupole components to appear. This step produces \(v_{i,\text{non-linear}}\). Finally, the linear and non-linear components are added to the mean velocities computed in step 1, leading to \(z_i=z_c + v_{i,\text{linear}}/c + v_{i,\text{non-linear}}/c\), with \(z_c\) computed in step 1.

    \item Using CF4 redshifts, we compute the distance modulus predicted by the $\Lambda$CDM model for each galaxy. To each computed distance, we add Gaussian noise using the reported uncertainties on $\mu$ from the CF4 catalog.
\end{enumerate}

These steps effectively produce a new `catalog'. In total, we have generated $100,000$ such datasets for each mask, repeating the pipeline described in Section~\ref{subsec:masks} for each simulation. Because it is numerically expensive to perform an MCMC analysis for each simulated dataset, for the steps above we only use the pseudo-inverse method. Thus, strictly speaking, these simulations are only meaningful for assessing the significance of the pseudo-inverse method outputs.

\section{Results}\label{sec:results}

We now report the results of our analysis using CF4 data and partial sky coverage. For simplicity, we only show the plots obtained with the \(\ns=8\) mask, since those obtained for \(\ns=4\) and 16 are qualitatively the same.

\subsection{Multipolar reconstruction}

\begin{figure}
    \centering
    \includegraphics[width=\linewidth]{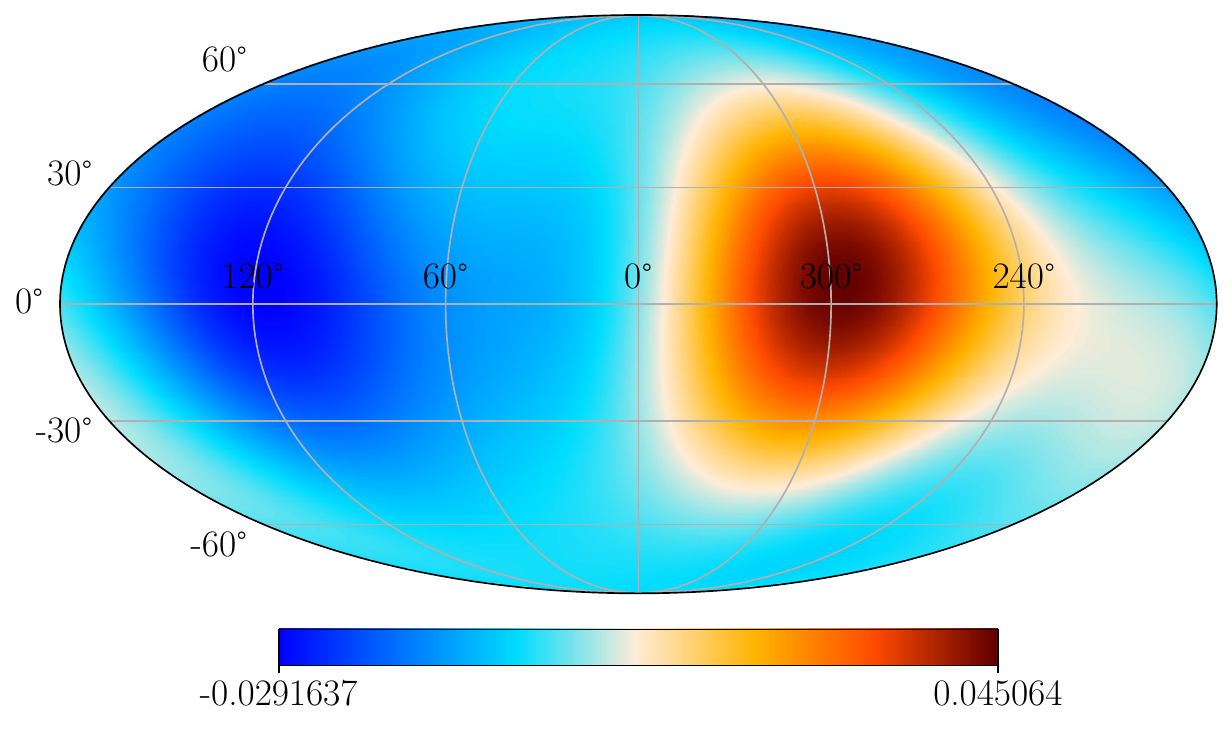}
    \caption{Fluctuation field $\eta$, reconstructed up to \(\ell=3\), using CF4 data and a mask with resolution $\ns=8$.}
    \label{fig:map_eta_reconstructed}
\end{figure}

The fluctuation field \(\eta\), reconstructed from a masked sky up to \(\ell=3\), is shown in Figure~\ref{fig:map_eta_reconstructed}. As we can see, this field is mainly dominated by a dipole with a maximum at galactic coordinates \((l,b)=(290^\circ,-4^\circ)\). The direction of this dipole is consistent with the bulk flow previously reported in the literature~\cite{Watkins:2008hf,Kalbouneh:2022tfw, Watkins:2023rll, hoffman2024large, kalbouneh2025anisotropic,Salzano:2025ang}.

Table \ref{tab:Cl_results} shows the reconstructed angular power spectrum of the fluctuation field for the first three multipoles. The mean values and error bars are computed directly from the posterior distribution of the coefficients \(\eta_{\ell m}\). We confirm earlier findings \cite{Kalbouneh:2022tfw}, based on CF3 data, showing that \(C_1\) is nearly an order of magnitude larger than \(C_2\) and \(C_3\), which is unexpected in a \(\Lambda\)CDM universe. We will revisit this question in Section \ref{subsec:alignments}. However, note that the values reported here are nearly three times larger than those based on CF3 data. This is probably due to the larger CF4 dataset, since we verified that the same masking procedure applied to CF3 data is consistent with the findings of \cite{Kalbouneh:2022tfw}.

\begin{table*}
    \begin{tabular}{c cc cc cc}
        \toprule
         & \multicolumn{2}{c}{$\ns=4$} & \multicolumn{2}{c}{$\ns=8$} & \multicolumn{2}{c}{$\ns=16$} \\
        \cmidrule(lr){2-3} \cmidrule(lr){4-5} \cmidrule(lr){6-7}
         & $C^\text{PI}_{\ell}$        & $C^\text{MCMC}_{\ell}$
         & $C^\text{PI}_{\ell}$        & $C^\text{MCMC}_{\ell}$
         & $C^\text{PI}_{\ell}$        & $C^\text{MCMC}_{\ell}$                                     \\

        \cmidrule(l{0.1em}r{0.1em}){1-1}
        \cmidrule(lr){2-3}\cmidrule(lr){4-5}\cmidrule(lr){6-7}

        $\ell=1$
         & $9.6 \pm 1.5$               & $9.7^{+1.9}_{-1.7}$
         & $9.9 \pm 1.4$               & $8.5^{+1.3}_{-1.2} $
         & $8.8 \pm 1.8$               & $9.0^{+1.5}_{-1.3}$                                        \\

        \cmidrule(l{0.1em}r{0.1em}){1-1}
        \cmidrule(lr){2-3}\cmidrule(lr){4-5}\cmidrule(lr){6-7}

        $\ell=2$
         & $0.44 \pm 0.17$             & $0.53^{+0.22}_{-0.19}$
         & $0.54 \pm 0.20$             & $0.53^{+0.21}_{-0.17}$
         & $0.88 \pm 0.30$             & $0.92^{+0.26}_{-0.22}$                                     \\

        \cmidrule(l{0.1em}r{0.1em}){1-1}
        \cmidrule(lr){2-3}\cmidrule(lr){4-5}\cmidrule(lr){6-7}

        $\ell=3$
         & $0.32 \pm 0.17$             & $0.57^{+0.24}_{-0.20}$
         & $0.46 \pm 0.20$             & $0.56^{+0.22}_{-0.18}$
         & $0.25 \pm 0.18$             & $0.40^{+0.17}_{-0.13}$                                     \\

        \bottomrule
    \end{tabular}
    \caption{Angular power spectrum \(C_{\ell}\) of the fluctuation field, reconstructed from the masked CF4 data using three different angular resolutions. All values are reported in units of \(10^{-4}\). Note that the results obtained from both the pseudo-inverse \((C_\ell^\text{PI})\) and MCMC \((C_\ell^\text{MCMC})\) reconstruction methods are in good agreement with each other across all multipoles and angular resolutions.}
    \label{tab:Cl_results}
\end{table*}

\begin{table}
    \begin{tabular}{c c c c}
        \toprule
         & $\ns=4$          & $\ns=8$          & $\ns=16$         \\
        \cmidrule(lr){2-2} \cmidrule(lr){3-3} \cmidrule(lr){4-4}
         & $(l,b)~(^\circ)$ & $(l,b)~(^\circ)$ & $(l,b)~(^\circ)$ \\

        \cmidrule(l{0.1em}r{0.1em}){1-1}
        \cmidrule(lr){2-2}\cmidrule(lr){3-3}\cmidrule(lr){4-4}

        $\ell=1$
         & $(290,-3)\pm6$
         & $(290,-4)\pm5$
         & $(289,-6)\pm5$                                         \\

        \cmidrule(l{0.1em}r{0.1em}){1-1}
        \cmidrule(lr){2-2}\cmidrule(lr){3-3}\cmidrule(lr){4-4}

        \multirow{2}{*}{$\ell=2$}
         & $(284,56)\pm22$
         & $(284,52)\pm19$
         & $(286,42)\pm14$                                         \\
         & $(329,-12)\pm17$
         & $(316,-19)\pm14$
         & $(332,-9)\pm11$                                        \\

        \cmidrule(l{0.1em}r{0.1em}){1-1} %
        \cmidrule(lr){2-2}\cmidrule(lr){3-3}\cmidrule(lr){4-4}

        \multirow{3}{*}{$\ell=3$}
         & $(262,-10)\pm14$
         & $(255,-15)\pm13$
         & $(261,-7)\pm13$                                       \\
         & $(325,-35)\pm15$
         & $(313,-27)\pm14$
         & $(339,-33)\pm14$                                       \\
         & $(340,54)\pm16$
         & $(336,47)\pm15$
         & $(344,60)\pm17$                                        \\

        \bottomrule
    \end{tabular}
    \caption{Galactic coordinates of the vectors \(\bv_{(1,1)}\), \(\bv_{(\alpha,2)}\), and \(\bv_{(\alpha,3)}\) of the fluctuation field, reconstructed from the masked CF4 data in the range \(z\in[0.01,0.05]\) with three different angular resolutions. The (symmetric) error bars correspond to the \(1\sigma\) Fréchet variance. These coordinates correspond to points in the hemisphere with north pole at \((290^\circ,-4^\circ)\); see Figure \ref{fig:mvs_final}.}
    \label{tab:mvs_final_results}
\end{table}

In order to compute the mean directions and the \(1\sigma\) uncertainties of the multipole vectors, we proceeded as follows: from the sampled posterior distribution of the coefficients \(\eta_{\ell m}\), as described in Section~\ref{subsec:MCMC}, we obtained the corresponding sample of multipole vectors. This leads, for each multipole \(\ell\), to  \(2\ell\) distinct clusters of points in the sphere, from which we want to derive the mean vector (i.e., the centroid of the cluster) and their \(1\sigma\) (Fréchet) standard deviations. However, because the multipole vectors do not have any natural ordering, it is not straightforward to delineate the boundaries of each cluster, particularly for the multipoles 2 and 3. To solve this problem, we resorted to the \texttt{k-means} clustering algorithm~\cite{scikit-learn} to separate the clusters. From each cluster we then computed the mean and \(1\sigma\) uncertainty using Eqs. \eqref{eq:frechet-mean} and \eqref{eq:frechet-var}. Our results are summarized in Table~\ref{tab:mvs_final_results} for each mask. The dipolar predominance of the signal is evident in the estimated error bars, since those of \(\ell=2\) and \(3\) are approximately three times larger than that of \(\ell=1\). We stress that the error bars reported in Table \ref{tab:mvs_final_results} have correlations which are not captured by our pipeline. Correlations arise both among vectors of the same multipole---since multipole vectors are not independently distributed, even in a Gaussian, homogeneous, and statistically isotropic universe~\cite{Dennis:2004sz}---and among vectors from different multipoles, since the full-sky reconstruction from partial-sky data is not exact, and mild correlations between different multipoles may still persist. A precise estimation of these correlations is not trivial, as the probability distribution of multipole vectors has compact support and a highly nontrivial structure \cite{Dennis:2004sz,Dennis:2007jk}. The quoted error bars should therefore be regarded as a first-order approximation.

\begin{figure}
    \includegraphics[width=\linewidth]{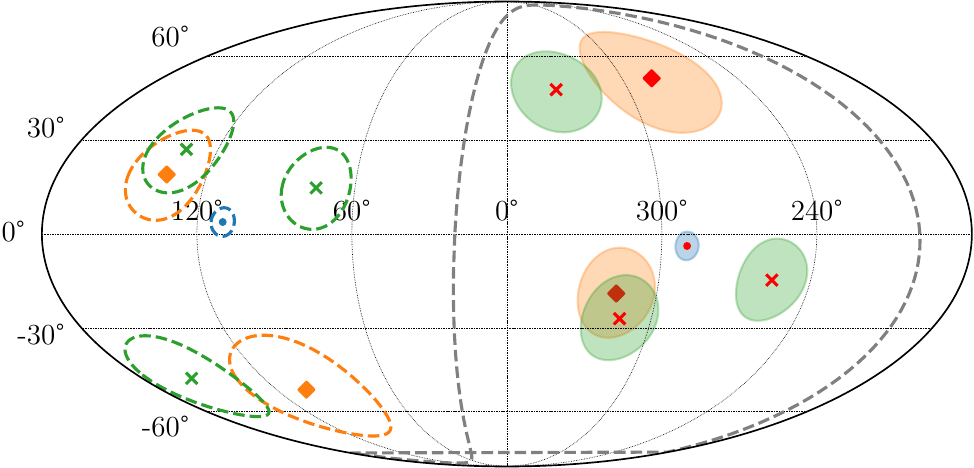}
    \caption{Reconstructed dipole (dots), quadrupole (diamonds) and octupole (crosses) directions and their Fréchet variances (blue, orange and green filled circles, respectively), for $\ns=8$. The dashed gray circle defines the equator having the dipole vector \(\bv_{(1,1)}\) as the \(z\)-axis. The points and dashed circles in the associated southern hemisphere correspond to the antipodes of each direction.}
    \label{fig:mvs_final}
\end{figure}

Finally, in Figure \ref{fig:mvs_final} we show the multipole vectors of the first three multipoles of the fluctuation field. As discussed in Sec.~\ref{subsec:mvs}, these vectors should be seen as axes crossing the sphere through the origin, since their overall orientations are not fixed by the data. Another important aspect of the STF formalism is that, in general, the axes \(\bv_{(\alpha,\ell)}\) do not agree with the location of the maxima and minima of the multipole \(\eta_\ell\) (see Figure 1 in \cite{Dennis:2007jk}). The only exception is the dipole vector \(\bv_{(1,1)}\), whose axis always aligns with the extrema of the dipole moment of the field.\footnote{In electrostatics, the dipolar electric potential is \(q\bv_{(1,1)}\cdot\boldsymbol{r}/r^3\), where \(\bv_{(1,1)}\) points from the negative to the positive charge.} Therefore, the directions of the vectors shown in Table~\ref{tab:mvs_final_results} and Figure \ref{fig:mvs_final} have been fixed as follows: first, we fixed the direction of \(\bv_{(1,1)}\) as the maximum of \(\eta_1\), which also aligns with the reported bulk flow. We then chose a hemisphere having this direction as the north pole, and fixed the directions of the remaining vectors as those pointing to this hemisphere. Our results are independent of this (arbitrary) prescription, and other conventions are possible.

\subsection{Likelihood of the dipole \(C_1\)}

As shown in Table~\ref{tab:Cl_results}, the dipole component of the angular power spectrum, \(C_1\), is one order of magnitude larger than \(C_2\) and \(C_3\). Since in a \(\Lambda\)CDM universe the fluctuation field is directly linked to the field of peculiar velocities (see Eq.~\eqref{eq:eta_v}), we can ask if these values are consistent with the theoretical predictions. In order to isolate the source of this effect, we implement the procedures described in Section~\ref{subsec:lcdm} for three different redshift bins: the full bin \(z\in[0.01,0.05]\), the left bin \(z\in[0.01,0.03]\) and the right bin \(z\in[0.03,0.05]\).

\begin{figure*}
    \centering
    \includegraphics[width=\linewidth]{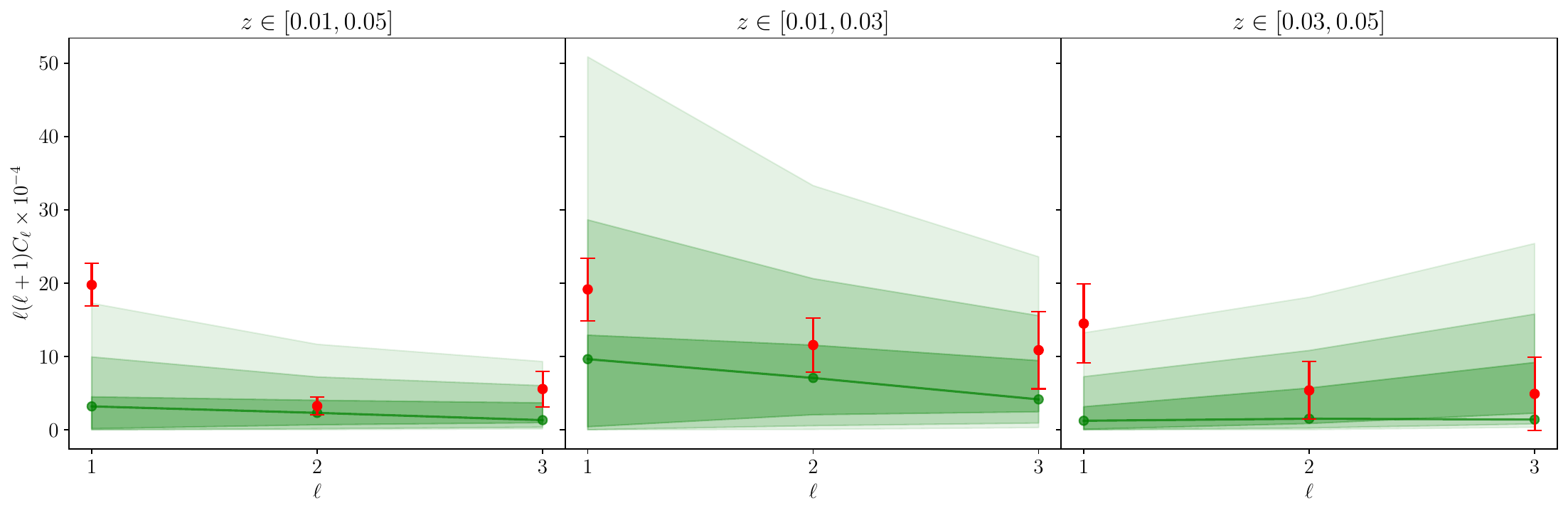}
    \caption{Comparison between the theoretical (green line) and measured (red dots) angular power spectrum using the pseudo-inverse method (see Section~\ref{subsec:lcdm}). The green regions represent the $68.3\%$, $95.4\%$ and $99.7\%$ C.L intervals computed from the distribution of the simulations using the \textit{Highest density interval} routine of the \texttt{arviz} \texttt{Python} package \cite{Martin2026}.}
    \label{fig:Comparison_all_z}
\end{figure*}

Our results are shown in Figure~\ref{fig:Comparison_all_z}. We first note that the confidence regions for both left and right bins are larger than that for the full bin. This is due to fewer data points in these bins, which increases the effects of errors and non-linearities. Moreover, in the left bin, the theoretical $C_{\ell}$ values are greater than the other cases, which is expected since peculiar velocities (and hence \(\eta\)) are larger at small redshifts. We also notice that in the right bin, the theoretical value of $C_3$ is below the $1\sigma$ interval of the simulations. This suggests that the octupole for this bin is dominated by noise.

Regarding the measured spectra, we notice that, for each of the three bins, the inferred quadrupole and octupole power spectra are within the $2\sigma$ interval of the simulations. The dipole, however, presents some tension with the theoretical expectations for the full (\(3.3\sigma\)) and right \((3.2\sigma)\) bins. In the left bin, the dipole is consistent with the $2\sigma$ interval of the expected theoretical value. This indicates that the source for the bulk flow is due to the more distant set of galaxies.\footnote{Note that this is also the regime most susceptible to Malmquist-type selection effects, which we have not separately modeled here.} Other works analyzing this dataset have found similar results \cite{Watkins:2023rll, whitford2023evaluating, hoffman2024large}.

\subsection{Testing for alignments between multipoles}\label{subsec:alignments}

\begin{table*}
    \centering
    \begin{tabular}{c cc cc cc}
        \toprule
         & \multicolumn{2}{c}{$N_{\text{side}}=4$}
         & \multicolumn{2}{c}{$N_{\text{side}}=8$}
         & \multicolumn{2}{c}{$N_{\text{side}}=16$}                    \\
        \cmidrule(lr){2-3} \cmidrule(lr){4-5} \cmidrule(lr){6-7}
        $(\ell,\ell')$
         & $S_{\ell,\ell'}$                         & $T_{\ell,\ell'}$
         & $S_{\ell,\ell'}$                         & $T_{\ell,\ell'}$
         & $S_{\ell,\ell'}$                         & $T_{\ell,\ell'}$ \\
        \cmidrule(l{0.5em}r{0.5em}){1-1}
        \cmidrule(lr){2-3} \cmidrule(lr){4-5} \cmidrule(lr){6-7}

        $(1,2)$
         & $0.52\,(0.84)$                           & $0.77\,(0.84)$
         & $0.50\,(0.87)$                           & $0.75\,(0.87)$
         & $0.46\,(0.94)$                           & $0.71\,(0.94)$   \\
        \cmidrule(l{0.5em}r{0.5em}){1-1}
        \cmidrule(lr){2-3} \cmidrule(lr){4-5} \cmidrule(lr){6-7}

        $(1,3)$
         & $0.37\,(0.56)$                           & $0.59\,(0.80)$
         & $0.35\,(0.45)$                           & $0.56\,(0.66)$
         & $0.39\,(0.70)$                           & $0.59\,(0.82)$   \\
        \cmidrule(l{0.5em}r{0.5em}){1-1}
        \cmidrule(lr){2-3} \cmidrule(lr){4-5} \cmidrule(lr){6-7}

        $(2,3)$
         & $0.54\,(0.36)$                           & $0.71\,(0.49)$
         & $0.55\,(0.32)$                           & $0.72\,(0.45)$
         & $0.51\,(0.47)$                           & $0.75\,(0.34)$   \\

        \bottomrule
    \end{tabular}
    \caption{Alignment statistics $S_{\ell,\ell'}$ and $T_{\ell,\ell'}$ for the first three multipole vectors derived from CF4 data. The values in parentheses correspond to the probabilities ($p$-values) derived from 10000 simulations.}\label{tab:ST_alignment}
\end{table*}

As a final task, we search for alignments between the multipoles of the fluctuation field. In Refs.~\cite{Kalbouneh:2022tfw,kalbouneh2025anisotropic}, the maxima of the first three multipoles of the fluctuation field were computed, and these directions were found to be nearly aligned. We also find that the maxima of these multipoles are close to the direction of the bulk flow. However, as discussed above, each multipole \(\eta_\ell\) is characterized by \(\ell\) unit vectors that generally do not coincide with its extrema. Since alignments are not expected to happen in a Gaussian, homogeneous, and isotropic universe, it is important to conduct a robust test of alignment based directly on the multipole vectors.

Several statistics sensitive to alignments between multipoles have been developed in the context of CMB low-multipole anomalies --- see \cite{abramo2006alignment} for a list. One way to circumvent the ambiguity in the orientation of the multipole vectors is to define statistics in terms of the ``area vectors'' \(\bw_{(\gamma,\ell)}\), normal to the planes formed by two multipole vectors:
\begin{equation}
    \boldsymbol{w}_{(\gamma, \ell)} = \boldsymbol{v}_{(\alpha, \ell)} \times \boldsymbol{v}_{(\beta, \ell)}\,,\qquad\gamma=1,\cdots,\lambda\,,
\end{equation}
where \(\lambda\equiv\ell(\ell-1)/2\) is the number of independent planes that can be constructed from each multipole \(\ell>1\). We then define the $S$ and $T$ statistics as
\begin{align}
    S_{\ell, \ell'} & = \frac{1}{\lambda\lambda'}\sum_{\gamma = 1}^{\lambda}\sum_{\gamma' = 1}^{\lambda'} \big|\boldsymbol{w}_{(\gamma, \ell)} \cdot \boldsymbol{w}_{(\gamma', \ell')}\big|\,,    \\
    T_{\ell,\ell'}  & = 1 - \frac{1}{\lambda\lambda'}\sum_{\gamma = 1}^{\lambda}\sum_{\gamma' = 1}^{\lambda'} (1 -|\boldsymbol{w}_{(\gamma, \ell)} \cdot \boldsymbol{w}_{(\gamma', \ell')}|)^2\,,
\end{align}
with the implicit assumption that \(\bw_{(1,1)}=\bv_{(1,1)}\). These quantities represent a straightforward generalization of those introduced in \cite{copi2006large, copi2015large}, in the sense that the definitions above allow testing alignments between multipoles \(\ell\) and \(\ell'\). It is easy to check that both statistics take values in the interval \([0,1]\).

To assess the likelihood of alignments observed in the dipole, quadrupole, and octupole vectors of CF4 data, we performed 10,000 simulations as described in the previous section, computing $S_{\ell,\ell'}$ and $T_{\ell,\ell'}$ for each simulation. We then extracted the two-tailed \(p\)-value. Our results are shown in Table \ref{tab:ST_alignment} for each of the masks we considered. As we can see, none of the computed alignments stand out as anomalous, with \(p\)-values not smaller (greater) than 32\% (94\%). For comparison, the same statistics applied to quadrupole and octupole multipole vectors of Planck data are below \(2\%\) \cite{copi2015large}. In fact, in the case of CMB, this alignment follows from the unusual planarity of the vectors \(\bv_{(\alpha,3)}\), along with the alignment of this plane with the plane formed by the vectors \(\bv_{(\alpha,2)}\) \cite{deOliveira-Costa:2003utu}. None of these features are present in CF4 data, as is evident in Figure \ref{fig:mvs_final}.

\section{Conclusions}\label{sec:conclusions}

The rich dataset of distances and redshifts provided by the Cosmicflows-4 catalog allows for a data-driven reconstruction of anisotropies in the local expansion rate of the universe, which is an important test of the Cosmological Principle. So far, the complete characterization of the angular degrees of freedom associated with each multipole of the local expansion rate has not been achieved. We have remedied this situation by means of a tensorial basis expansion, where each multipole \(\ell\) of the expansion rate is characterized by one amplitude and \(\ell\) unit (multipole) vectors, totalizing the expected \(2\ell+1\) real degrees of freedom per multipole. This basis is particularly suited for analysis of directionalities, since these vectors rotate rigidly with the data, making the whole statistical pipeline less susceptible to external choices of frames.

To mitigate spurious anisotropies arising from non-uniform angular and radial distributions of distances and redshifts in the catalog, we employed a pixel-based masking procedure that accounts for sparsely populated or empty regions in the sky. The full-sky expansion rate fluctuation field was then reconstructed using two complementary methods: one based on a pseudo-inverse of the multipole-coupling kernel, and an independent Bayesian analysis.

We confirm the existence of a strong dipole in the direction \((l,b)=(290^\circ,-4^\circ)\pm5^\circ\), which agrees with previous reports of a coherent field of peculiar velocities, or bulk flow. The two vectors associated with the quadrupole and the three vectors associated with the octupole show no indication of alignments between them or with the dipole. The quadrupole and octupole vectors also show no apparent connection with the axes of the equivalent CMB multipole vectors. Nevertheless, it is intriguing that the maxima of the moments \(\eta_1\), \(\eta_2\), and \(\eta_3\) are nearly aligned with the bulk-flow direction, and a dedicated analysis will be necessary to determine whether this feature reflects residual systematics or a genuine physical effect. This is left as a future investigation.

We have found that the angular power spectrum of the local expansion rate reconstructed from CF4 data is nearly three times larger than the same signal obtained from CF3 data. However, we have found that the dipole \(C_1\) is one order of magnitude larger than the quadrupole \(C_2\) and octupole \(C_3\), for both datasets. To quantify this disagreement with predictions from a base \(\Lambda\)CDM model, we derived the theoretical angular power spectrum for the expansion rate fluctuation field \(\eta\) and its statistical and systematic dispersion. This allowed us to derive a \(3.3\sigma\) tension between the measured \(C_1\) and \(\Lambda\)CDM predictions. Moreover, we found that this tension is mainly due to sources in \(z\in[0.03,0.05]\). A more detailed tomographic analysis, combined with independent large-scale tracers, will be crucial to determine whether this discrepancy reflects residual systematics --- such as Malmquist bias or calibration offsets between CF4's distance-indicator subsamples --- or a genuine deviation from the \(\Lambda\)CDM model.

\begin{acknowledgements}
    J. G. V is supported by Conselho Nacional de Desenvolvimento Científico e Tecnológico (CNPq). T.S.P and S.D.P.V are supported by Fundação Araucária (NAPI Fenômenos Extremos do Universo, grant 347/2024 PD\&I). R. R. G and V. M. G are supported by Coordenação de Aperfeiçoamento de Pessoal de Nível Superior (CAPES).
\end{acknowledgements}

\appendix

\section{Derivation of Eq. \eqref{eq:Cl_lcdm}}\label{sec:Cl_lcdm}

In this section we derive the angular power spectrum of the fluctuation field \(\eta\), related to the peculiar velocity \(v\) through Eq.~\eqref{eq:eta_v}. We begin by connecting the velocity field to the matter density contrast. On sub-horizon scales and at low redshifts, the longitudinal velocity \(\boldv\) is related to the density contrast \(\delta\) through the continuity equation
\begin{equation*}
    \boldv(\boldk,\chi)\simeq i\,\frac{\boldk}{k^2}\,aHf\,\delta(\boldk,\chi),
\end{equation*}
where $\chi$ is the comoving distance and \(f\) the linear growth rate. We are interested in the line-of-sight component $v\equiv\hn\cdot\boldv$, so that the fluctuation field can be written as
\begin{equation*}
    \eta(\boldx,\chi)=\frac{iaHf}{cz\ln 10}\int\frac{\dd^3k}{(2\pi)^3}\frac{\hn\cdot\boldk}{k^2}e^{i\boldk\cdot\boldx}\delta(\boldk,\chi)\,.
\end{equation*}

The two-point correlation function of $\eta$ now follows from that of the density contrast,
\begin{equation*}
    \langle\delta(\boldk,\chi)\delta(\boldk',\chi')\rangle
    =(2\pi)^3 P_m(k,\chi,\chi')\,\delta^{3}(\boldk+\boldk')\,,
\end{equation*}
where \(P_m\) is the matter power spectrum. Setting $\boldx=\chi\hn$, $\boldx'=\chi'\hn'$, and defining $\mu=\hk\cdot\hn$, $\mu'=\hk\cdot\hn'$, we obtain
\begin{align*}
    \langle\eta(\boldx,\chi)\eta(\boldx',\chi')\rangle
    = & \frac{A(\chi)A(\chi')}{(\ln 10)^2}
    \int\frac{k^2\dd k}{(2\pi)^3}\frac{P_m(k,\chi,\chi')}{k^2} \nonumber \\
      & \times \frac{1}{k^2}\frac{\dd}{\dd\chi}\frac{\dd}{\dd\chi'}
    \int\dd^2\hk\,e^{ik(\mu\chi-\mu'\chi')},
\end{align*}
where, for simplicity, we have defined $A\equiv aHf/(cz)$.

The mathematical identities
\begin{align*}
    e^{ik\mu\chi} & = \sum_{\ell=0}^{\infty} i^\ell j_\ell(k\chi)P_\ell(\mu),                           \\
    P_\ell(\hn\cdot\hn')\delta_{\ell\ell'}
                  & = \frac{2\ell+1}{4\pi}\int \dd^2\hk\, P_\ell(\hk\cdot\hn)P_{\ell'}(\hk\cdot\hn')\,,
\end{align*}
allow us to express the correlation function as
\begin{equation*}
    \langle\eta(\boldx,\chi)\eta(\boldx',\chi')\rangle
    = \sum_{\ell}\frac{2\ell+1}{4\pi}
    C_\ell(\chi,\chi')P_\ell(\hn\cdot\hn'),
\end{equation*}
where the time-dependent power spectrum given by
\begin{equation*}
    C_\ell(\chi,\chi')=
    \frac{A(\chi)A(\chi')}{(\ln 10)^2}\frac{2}{\pi}
    \int \dd k\,P_m(k,\chi,\chi')\,
    j'_\ell(k\chi)\,j'_\ell(k\chi')\,.
\end{equation*}
Here, a prime in \(j_\ell\) denotes differentiation with respect to its argument.

To obtain a time-independent angular power spectrum, we average over conformal time with a selection function ${\cal W}(\chi)$. For simplicity, we adopt a normalized top-hat window,
\begin{equation*}
    {\cal W}(\chi)=
    \begin{cases}
        \frac{1}{\Delta\chi} & \text{if } \chi_{\text{min}}\le\chi\le\chi_{\text{max}}, \\
        0                    & \text{otherwise}\,.
    \end{cases}
\end{equation*}

Over the narrow redshift range considered, $a$, $H$, and $f$ vary slowly compared to the oscillatory behavior of $j'_\ell$. To proceed, we use the geometric-mean approximation for the unequal-time power spectrum, \(P_m(k,\chi,\chi')\simeq \sqrt{P_m(k,\chi)}\sqrt{P_m(k,\chi')}\) \cite{Kitching:2016xcl}, as well as the approximations \(z\simeq H_0\chi/c\), \(a\simeq a_0\) and \(H\simeq H_0\). This finally gives
\begin{equation*}
    C_\ell=
    \frac{2f^2}{\pi(\ln 10)^2\Delta\chi}
    \int \dd k\,P_m(k)
    \left(\int_{\chi_{\text{min}}}^{\chi_{\text{max}}}
    \dd\chi\,\frac{j'_\ell(k\chi)}{\chi}\right)^2.
\end{equation*}
Note that all factors of \(c\) cancel, so that the final expression for \(C_\ell\) is independent of \(c\), as it must be for the power spectrum of the dimensionless field \(\eta\).

\bibliographystyle{h-physrev4}
\bibliography{hubblemultipoles}

\end{document}